\documentclass[conference]{IEEEtran}
\usepackage[utf8]{inputenc}
\usepackage{listings}
\usepackage{comment}
\usepackage{graphicx}
\usepackage{color}
\usepackage{url}
\usepackage{booktabs}
\usepackage{multirow}
\usepackage{cite}
\usepackage[table,xcdraw]{xcolor}
\usepackage{subfig}
\usepackage{hyperref}
\usepackage{amsmath, amssymb}

\newcommand{\autohls}{{AutoHLS}\hspace{2pt}}

\title{\autohls: Learning to Accelerate Design Space Exploration for HLS Designs}

\author{
\IEEEauthorblockN{Md Rubel Ahmed, Toshiaki Koike-Akino, Kieran Parsons, Ye Wang}
Mitsubishi Electric Research Laboratories (MERL), 201 Broadway, Cambridge, MA 02139, USA.\\
\{mdahmed, koike, parsons, yewang\}@merl.com}
\date{November 2022}

\begin{document}

\maketitle

\begin{abstract}
High-level synthesis (HLS) is a design flow that leverages modern language features and flexibility, such as complex data structures, inheritance, templates, etc., to prototype hardware designs rapidly. 
However, exploring various design space parameters can take much time and effort for hardware engineers to meet specific design specifications. 
This paper proposes a novel framework called AutoHLS, which integrates a deep neural network (DNN) with Bayesian optimization (BO) to accelerate HLS hardware design optimization. 
Our tool focuses on HLS pragma exploration and operation transformation. 
It utilizes integrated DNNs to predict synthesizability within a given FPGA resource budget. 
We also investigate the potential of emerging quantum neural networks (QNNs) instead of classical DNNs for the AutoHLS pipeline. 
Our experimental results demonstrate up to a 70-fold speedup in exploration time.

\end{abstract}


\section{Introduction}
\label{motivation}

HLS is a widely used rapid design and prototyping method in industry and academia. Still, it poses several challenges for source code optimization due to the rich features of modern programming languages such as C/C++. Careless optimization can result in inefficient and resource-hungry designs with high latency or, in some cases, loss of synthesizability under a reasonable FPGA resource budget. HLS compilers such as Vitis~\cite{vitis_hls} offer optimization tactics such as pragma directives and timing/closure analysis to tackle these issues which have spurred active research areas in design-space exploration (DSE) for HLS. Accelerated DSE is required since downstream tools used for RTL generation, such as Vitis~\cite{vitis_hls}, can take significant time to compile and report synthesis results. This limits the number of designs evaluated during DSE, resulting in sub-optimal solutions. Besides, the time required for RTL generation can increase the DSE time from hours to days, depending on the complexity of the design. The quest for faster and more efficient DSE in HLS has led to the development of machine learning (ML) and analytical methods. In this context, ScaleHLS~\cite{scales} presents an analytical approach that leverages a Quality-of-Results (QoR) estimator to accelerate the DSE process. By statically analyzing code blocks and modeling latency and resource utilization, the QoR estimator enables ScaleHLS's DSE engine to explore the design space efficiently and converge to the Pareto front faster. Other methods~\cite{ml_for_eda_2021, scales, GNN, autoScaleDSE, mwscas, fahim2021hls4ml} use statistical, heuristic, ML, or meta-learning approaches to accelerate DSE. For instance, using an ML model, Pyramid~\cite{pyramid_clock_estimate} estimates the maximum achievable throughput. At the same time, a recent work~\cite{robust_estimation} predicts resource usage for synthesizing convolutional neural networks. Sherlock~\cite{sherlock2022}, another DSE tool, uses active learning with a surrogate model to find Pareto front, highlighting the challenges in handling conflicting objectives in parameter optimization. We consider Optuna~\cite{optuna_2019}, a Bayesian Optimization (BO) framework, as a baseline multi-objective optimization tool. BO is generally slow to find the Pareto front as the downstream HLS flow takes much time to generate QoR for each sample design point. Therefore, we add an early failure prediction network with the BO to accelerate the DSE. To the best of our knowledge, no current works focus on reducing the search space based on synthesizability constraints, such as FPGA footprints (DSP, FF, LUT) or synthesis time budget.

Our proposed method, \autohls, optimizes the design by considering synthesizability constraints as a multi-objective optimization problem. AutoHLS efficiently determine loop unrolling factor, pipeline depth, array partition, etc., for pragma installments in order to optimize HLS designs considering signal processor (DSP), flip-flop (FF), look-up table (LUT), power consumption, and latency. Furthermore, AutoHLS also includes kernel operations transformation to further optimize the designs. The {\bf contributions} of this work are as follows.

\begin{figure}[t]
 \centering
 \subfloat[regular multiplication]{
  \includegraphics[width=.4\linewidth]{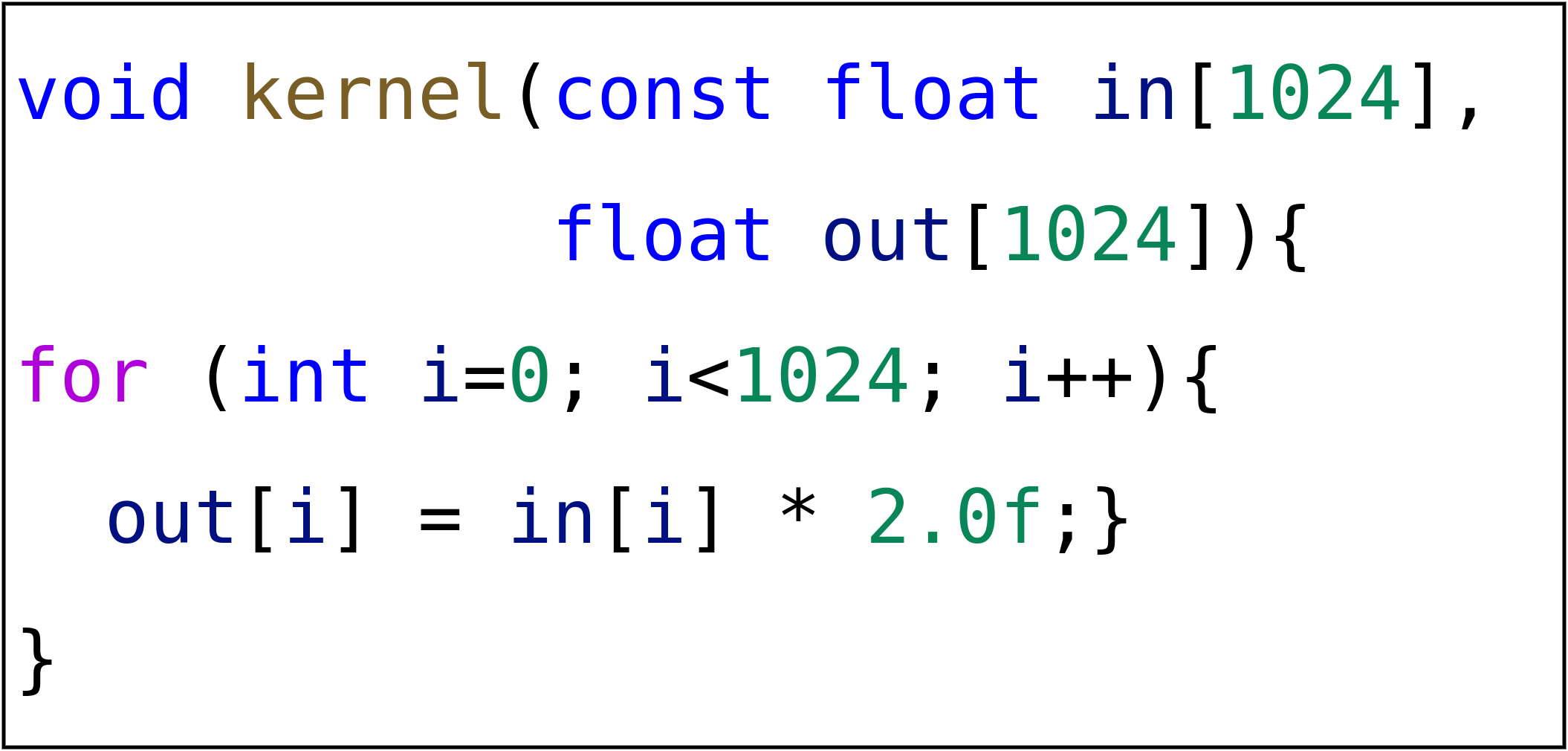}
 }
 \subfloat[exponent addition]{
 \includegraphics[width=.57\linewidth]{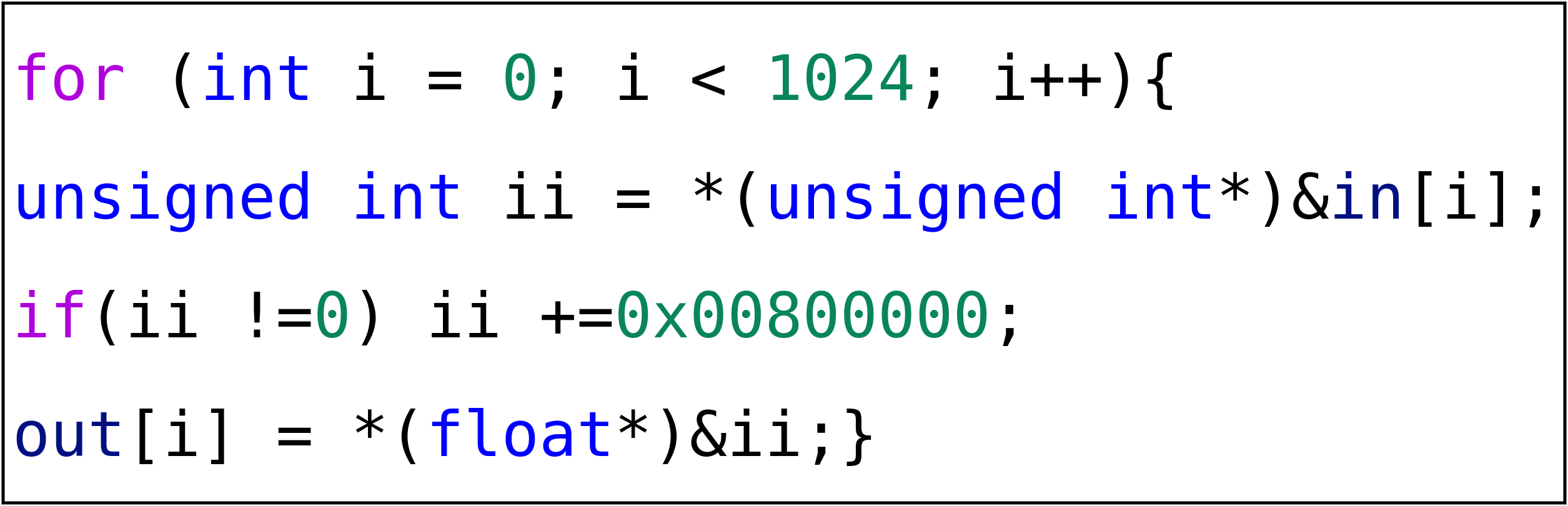}
 }
 \\
 \subfloat[pragma insertion]{
  \includegraphics[width=.55\linewidth]{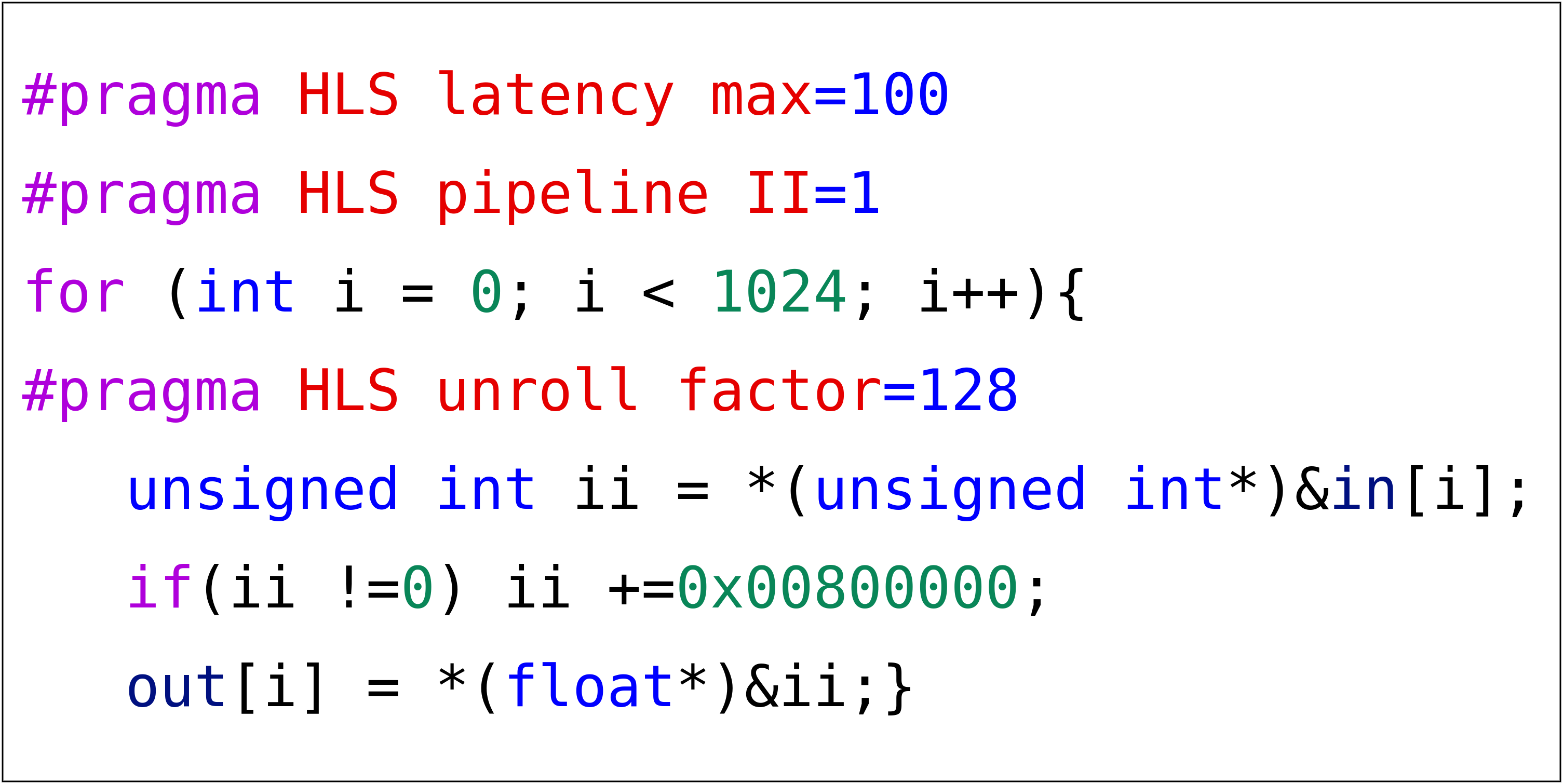} 
 }
 \subfloat[profiling result]{
  \includegraphics[width=.42\linewidth]{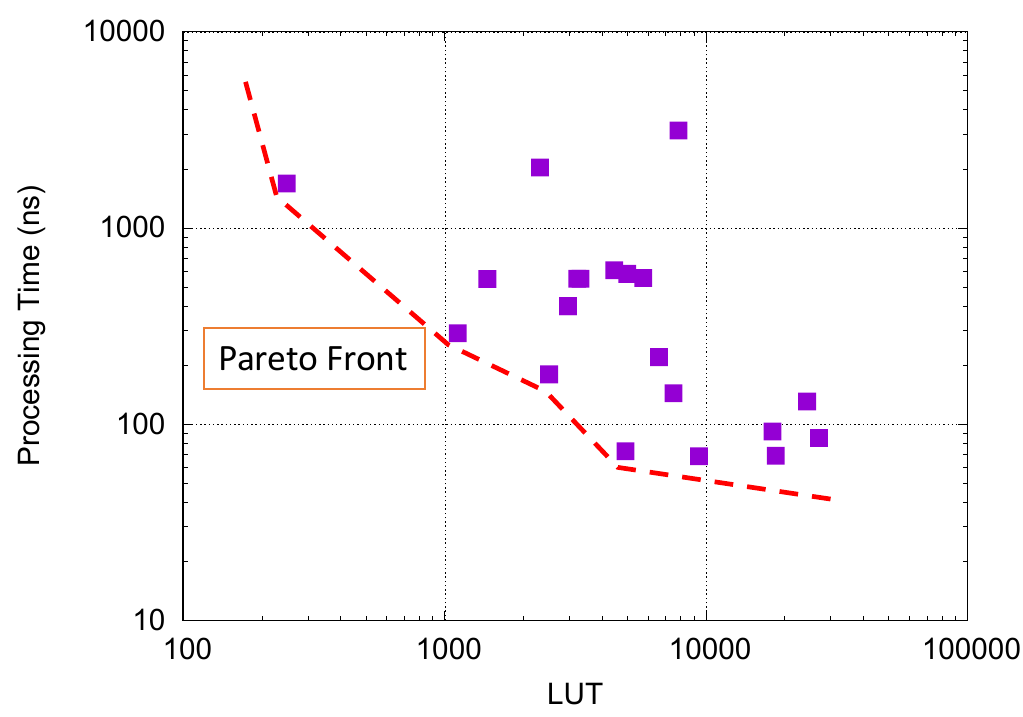} 
 }
 \caption{Array doubling kernels having functionally equivalent operations but different hardware profiles.}
 \vspace{-15pt}
 \label{fig:doubling}
\end{figure}

\begin{itemize}
    \item We reveal that existing multi-objective optimization tools can fail to meet the budget-centric design approach.
    \item We propose \autohls framework to accelerate the DSE using ML models.
    \item A novel QNN model is employed to predict synthesis failure and resource usage accurately.
\end{itemize}

\section{AutoHLS for Efficient Hardware Design}
\label{sec:autohls}
\begin{figure}[t]
\centering
\includegraphics[width=0.84\linewidth]{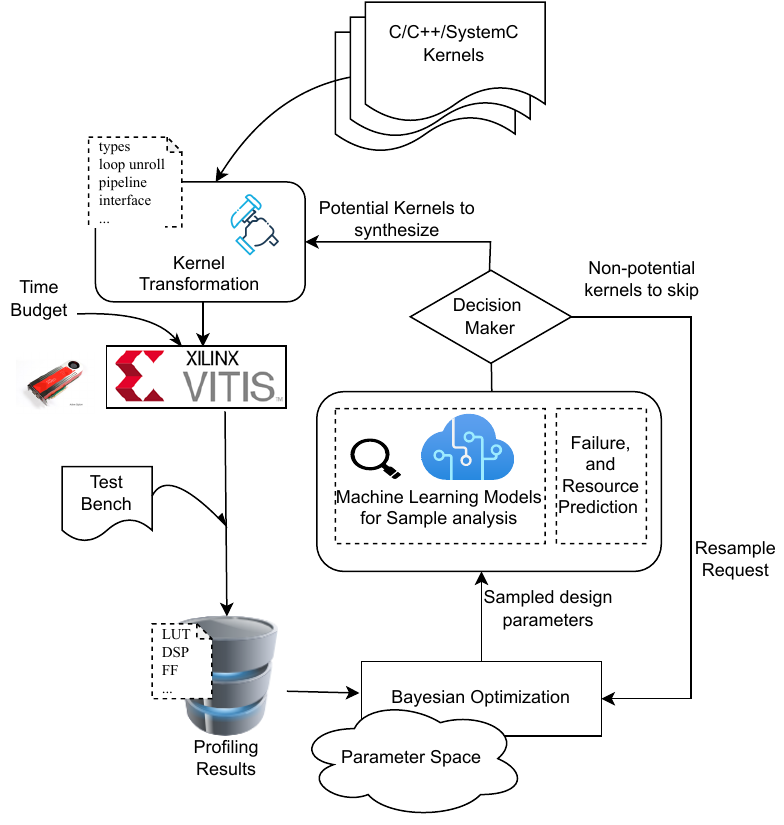}
\caption{Overview of AutoHLS.}
\vspace{-15pt}
\label{fig:autohls}
\end{figure}

In designing hardware in HLS, several critical factors must be considered, such as the target device, available resources, required precision level, simulation, synthesis, co-simulation time, etc. In this regard, Fig.~\ref{fig:doubling} (a) presents a regular array doubling kernel in C++ as an example. However, HLS provides several alternative implementations, such as Fig.~\ref{fig:doubling} (b), where a functionally equivalent exponent addition replaces the multiplication operation. Additionally, Fig.~\ref{fig:doubling} (c) shows a more optimized implementation that utilizes pragma insertion. The synthesis profiling results over different pragma factors and kernel operation transforms, shown in Fig.~\ref{fig:doubling} (d), demonstrate a tradeoff behavior in multi-objective optimization, where the reduction in LUT resources and runtime would compete. Nevertheless, due to the high degree of flexibility in pragma installment and kernel operation transforms, finding an optimal Pareto front in constrained development time remains challenging. \autohls, as depicted in Fig.~\ref{fig:autohls}, takes an unoptimized kernel and efficiently explores different design alternatives to meet design objectives such as runtime, precision level, DSP, FF, LUT, etc., usage. It can discover an optimal set of pragma and kernel operation transforms with the help of an ML-based synthesizability prediction mechanism.

\subsection{Scope and Definition}

\subsubsection{Pragma Selection}
Pragma and their parameters guide the HLS compiler toward optimal designs. For example, AutoHLS uses a categorical sampling of BO to decide the set of HLS pragma insertions ${P_K} \subseteq \mathbf{P}$, where $\mathbf{P}$ includes {\tt pipeline}, {\tt unroll}, etc.

\subsubsection{Pragma Parameter Selection}
Each HLS pragma $P$ can have a set of parameters $\mathbf{A}_P$. 
Given a kernel $K$, AutoHLS decides a parameter set $A_K \subseteq \{\mathbf{A}_P\}$ for each HLS pragma $P\in P_K$ in the selection, using BO sampling.
For example, Fig.~\ref{fig:doubling} (c) uses the parameters set of $A_K=\{100, 1, 128\}$ for the pragma set $P_K$.

\subsubsection{Kernel/Operation Transformation}
HLS synthesis tools often utilize high-cost resources, such as DSP blocks, to meet high throughput requirements, which may not be available for resource-constrained applications like edge/embedded devices. Therefore, considering alternative operations that can save resources at a potential cost of throughput or precision. For example, a regular multiplication kernel in Fig.\ref{fig:doubling} (a) can be functionally equivalent to an exponent addition kernel in Fig.\ref{fig:doubling} (b) for a floating-point operation when the multiplicand is a power-of-two (PoT) value. Furthermore, simplifications can be achieved by reducing bit-width precision and using fixed-point operations with bit-shifting.
Given an HLS kernel $K$, \emph{kernel/operation transformation} produces another kernel ${K_T}$ such that the outputs from both kernels are almost equivalent or exactly equivalent within a specified tolerance range. In addition, recent green ML models have also demonstrated that quantized DNNs, such as DeepShift~\cite{pot_toshi}, can outperform floating-point DNNs. Therefore, we explore PoT and additive-Power-of-Two (APoT) quantization for further optimization.

\subsection{AutoHLS Flow}
\autohls explores both kernel and parameter space. Given a set of kernels ${K}$, an objective function, and an HLS design constraint, AutoHLS analyzes the kernels and returns a set of optimal synthesizable kernels for the given objectives that meet the design constraint.

\subsubsection{Kernel Transformation}
AutoHLS first parses the input C/C++ kernels and constructs pragmas using the selected set $\mathbf{P}$, which includes the pipeline, unroll, latency, array partition, etc. These kernels are then checked for feasibility before being synthesized.

\subsubsection{Kernel Synthesis}
The transformed kernel is synthesized using standard HLS tools (e.g., Vitis) with pre-set device-specific parameters for FPGA. The synthesis process involves functional correctness checking with {\tt csim} and feasibility checking with {\tt synth}.

\subsubsection{Kernel Profiling}
After the synthesis step, the Quality of Results (QoR), kernel type, and pragma parameters are collected. The synthesis can be complete or fail for the given constraint. These data are utilized directly or indirectly in the objective function.

\subsubsection{Bayesian Optimization}
\autohls adopts the BO method based on a tree-structured Parzen estimator (TPE)~\cite{optuna_2019} for DSE, which can handle multi-objective optimization. The TPE-based optimizer suggests a set of optimized design parameters from the parameter space based on an acquisition function for efficient Pareto optimization.

\subsubsection{Decision Maker}
AutoHLS tool incorporates machine learning techniques to predict the synthesis failure and estimate the resource utilization of the designed kernel. Specifically, DNN and QNN provide the failure prediction scores on each sample set generated by the BO. Based on the prediction results, the tool decides whether to synthesize or discard the kernel and move to the next one. This approach enables accelerated design space exploration and reduces the overall design time.

\subsection{ML models for Sample Analysis}
AutoHLS employs ML models to predict synthesis failure and estimate the resource profile of a design. These models, including classifiers and regression models, are trained on the already explored samples and assign a score to a new sample generated by BO. A decision is then made based on a threshold $\tau$. Finally, the sample is sent for synthesis only if it passes the decision-maker.

\begin{figure}[t]
\centering
\subfloat[][DNN]{
\includegraphics[width=.45\linewidth]{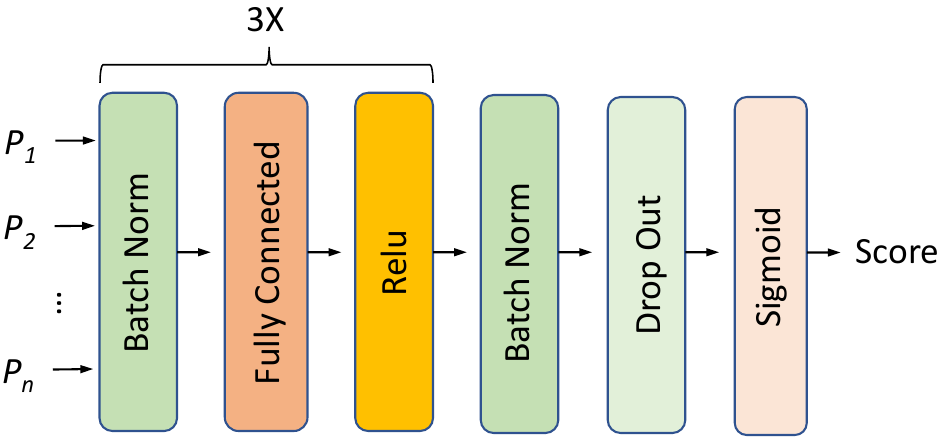}}
\subfloat[][Variational QNN]{
\includegraphics[width=0.53\linewidth]{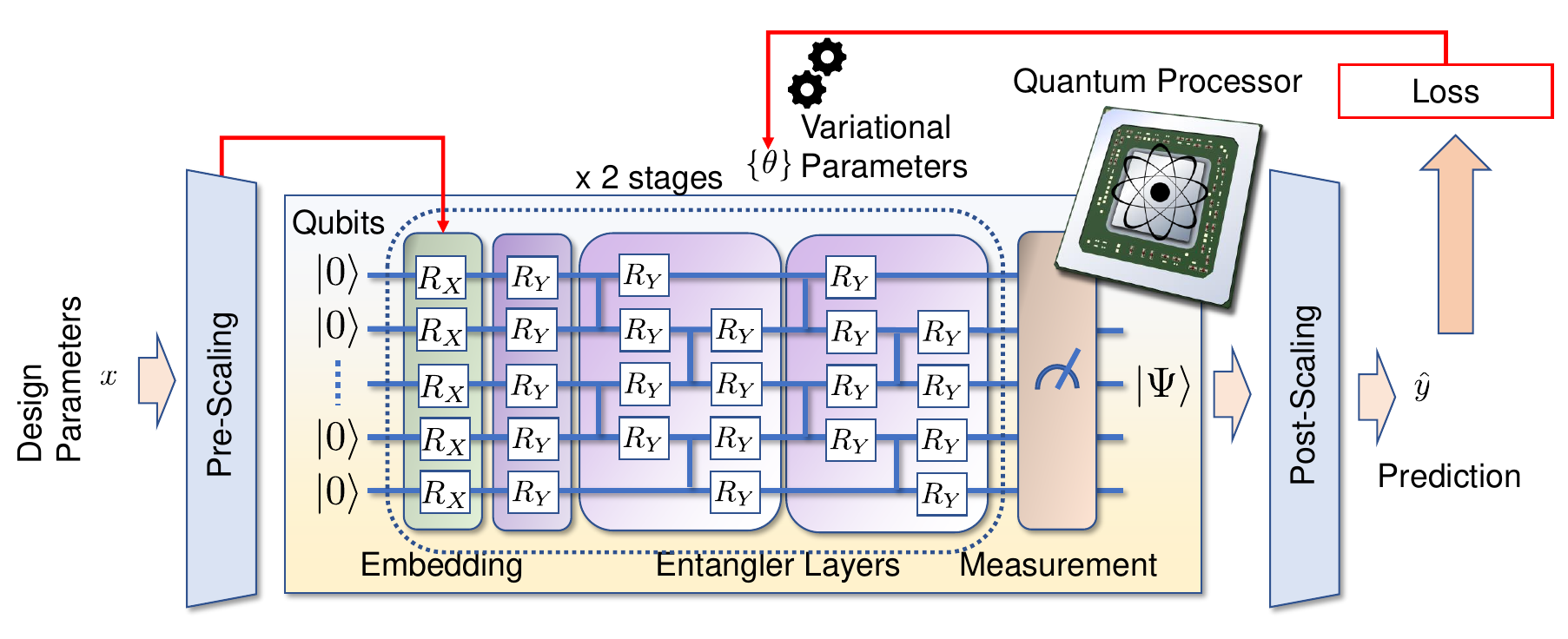}}
\caption{Failure/resource prediction models.}
\vspace{-15pt}
\label{fig:dnn-qnn}
\end{figure}

\subsubsection{DNN}
We propose a DNN model shown in Fig.~\ref{fig:dnn-qnn} (a), for predicting a design's synthesizability score and resource usage. The model takes design parameters as input and consists of three batch normalization layers, a fully-connected layer, and a Relu activation. A batch normalization layer, a dropout, and a sigmoid are applied at the end. The model has 3243 trainable parameters and is designed to learn with limited training samples, which is essential in DSE due to the long synthesis time of HLS tools.

\subsubsection{QNN}
The recent advancements in quantum technology have led to the availability of high-qubit processors, such as the 433-qubit processors released by IBM in 2022. This has given rise to a new paradigm of ML models known as QNNs, which have the universal approximation property~\cite{PerezSalinas2020datareuploading} and are more compact than modern DNNs. We propose a proof-of-concept evaluation of QNNs for HLS acceleration. First, we present a QNN architecture shown in Fig.~\ref{fig:dnn-qnn} (b) with only 54 trainable parameters. It has five quantum bits.

\subsubsection{Classical ML Algorithms}
We evaluate various classical ML models, including SVM and LR for failure prediction and linear regression, lasso, KRR, and Bayesian ridge regression for hardware profile prediction. 
\section{AutoHLS Validation}
\label{sec:eval}
Our experiments are performed on a machine with an Intel{\textregistered} Core{\texttrademark} i7-8700K CPU @ 3.70GHz and 64GB of main memory, running on Ubuntu 20.04.5 LTS. The Xilinx ZCU104 board is used as a target FPGA, and Vitis HLS 2022.1 is used for kernel synthesis.

\subsection{Problem Setup}
We investigate the effectiveness of AutoHLS for the DSE of a CNN block. We consider synthesis time $t$ as a design resource budget or constraint. The CNN block comprises a window size $L$, an input channel $C_{\mathrm{in}}$, and an output channel $C_{\mathrm{out}}$, where the convolution operation involves element-wise multiplication and accumulation of the window and input channel elements. Table~\ref{tab:mac_transforms} provides the area utilization of the conventional multiplier-based implementation. MAC stands for multiplication and accumulation-based convolution. In Table~\ref{tab:mac_transforms}, we present the QoR results for $C_{\mathrm{in}} = 100$, $L = 7$, $C_{\mathrm{out}} = 106$, and $\tt float32$ as the datatype. The table shows different types of kernels, such as PoT\textless{}16, 6\textgreater{} and APoT\textless{}16, 6\textgreater{}, which are arbitrary precision (ap) fixed-point data types. The results reveal kernel transformation significantly impacts the hardware footprint. However, kernel transformation may cause some loss in precision, which MSE indicates. Additionally, the ap-type PoT has the lowest resource usage but higher MSE than APoT, which has better MSE but consumes more area than PoT. Finding the optimal kernel requires DSE to determine the pragma and appropriate pragma parameters. We conduct a case study on two kernels, PoT and APoT that entirely eliminates the multipliers. We then evaluate the performance of the conventional BO method and subsequently employ \autohls for further optimization.

\subsubsection{Quantizations}
We use PoT and APoT quantizations as kernel transformation schemes to create hardware-friendly designs of green ML models~\cite{pot_toshi}.
A regular MAC with $W$ as a weight, $b$ as the bias:
$y = Wx + b$;
the PoT quantization of weight, $W$, $u \in \mathbb{Z}$:
$W = \pm 2^{u}$;
and APoT quantization of weight, $W$:
$W = \pm 2^{u} \pm 2^{v}$,
where $u, v \in \mathbb{Z}$ and $v<u$.

\subsubsection{Bayesian Optimization for DSE}

To investigate the performance of BO on parameter optimization, the kernels are instrumented with four pragmas: unroll factor, pipeline instantiation interval, latency max and min. 

\begin{table}[!t]
\caption{Resource usage of convolution kernels}
\label{tab:mac_transforms}
\centering
\begin{tabular}{llllll}
\toprule
\textbf{Kernel} &
  \multicolumn{1}{c}{\textbf{FF}} &
  \multicolumn{1}{c}{\textbf{LUT}} &
  \multicolumn{1}{c}{\textbf{DSP}} &
  \textbf{Latency} &
  \multicolumn{1}{c}{\textbf{MSE}} \\ \midrule
MAC                               & 40922 & 17761 & 5 & 3072 & -                                      \\ 
MAC\textless{}16, 6\textgreater{}  & 24784 & 8650 & 1 & 1352 & 5.09e-06                                      \\ 
PoT                               & 42396 & 18067 & 4 & 4533 & 3.78                                   \\ 
PoT\textless{}16, 6\textgreater{} &
  24893 &
  6947 &
  \cellcolor[HTML]{FFFFFF}\textbf{0} &
  \cellcolor[HTML]{FFFFFF}\textbf{925} &
  3.78 \\ 
APoT                              & 45207 & 18593 & 4 & 4952 & 0.019                                  \\ 
APoT\textless{}16, 6\textgreater{} & 26061 & 7090  & 0 & 1039 & \cellcolor[HTML]{FFFFFF}\textbf{0.019} \\ \bottomrule
\end{tabular}
\vspace{-15pt}
\end{table}

Table~\ref{tab:data-gen} presents the performance evaluation of BO on the exploration process. The column `Time' indicates the synthesis time budget in minutes. The columns `Comp.' and `Fail' denote the number of samples for which the kernel synthesis succeeded and failed, respectively. The exploration involves $3302$ designs, taking $4$ to $6$ minutes to complete, regardless of the synthesis status. However, most of the parameters suggested by BO failed to synthesize, resulting in a vain attempt to synthesize the wrong design. To address this problem, \autohls leverages an early failure prediction mechanism.

\begin{table}[!t]
\caption{Kernel synthesis using BO with a given time budget}
\label{tab:data-gen}
\centering
\begin{tabular}{@{}lllllll@{}}
\toprule
\multicolumn{1}{c}{\textbf{\hspace{-7pt}Kernel}} & \multicolumn{1}{c}{\textbf{\begin{tabular}[c]{@{}c@{}}Time\\ (min.)\end{tabular}}} & \multicolumn{1}{c}{\textbf{Comp.}} & \multicolumn{1}{c}{\textbf{Fail}} & \multicolumn{1}{c}{\textbf{\begin{tabular}[c]{@{}c@{}}Comp. \\ + Fail\end{tabular}}} & \multicolumn{1}{c}{\textbf{\%Fail}} & \multicolumn{1}{c}{\textbf{\%Comp.}} \\ \midrule
\multirow{5}{*}{\textbf{APoT}}      & 2.00                                                                               & 14                                 & 194                               & 208                                                                                  & \textbf{93.26}                      & 6.73                                 \\
                                    & 2.20                                                                               & 12                                 & 289                               & 301                                                                                  & \textbf{96.01}                      & 3.98                                 \\
                                    & 2.50                                                                               & 20                                 & 480                               & 500                                                                                  & \textbf{96.00}                      & 4.00                                 \\
                                    & 2.75                                                                               & 11                                 & 389                               & 400                                                                                  & \textbf{97.25}                      & 2.75                                 \\
                                    & 3.00                                                                               & 42                                 & 551                               & 593                                                                                  & \textbf{92.76}                      & 7.07                                 \\ \midrule
\multirow{4}{*}{\textbf{PoT}}       & 1.50                                                                               & 19                                 & 381                               & 400                                                                                  & \textbf{95.25}                      & 4.75                                 \\
                                    & 1.75                                                                               & 364                                & 36                                & 400                                                                                  & 9.00                                & 94.00                                \\
                                    & 2.00                                                                               & 291                                & 9                                 & 300                                                                                  & 3.00                                & 97.00                                \\
                                    & 2.20                                                                               & 188                                & 12                                & 200                                                                                  & 6.00                                & 94.00                                \\ \midrule
\multicolumn{2}{l}{\hspace{-7pt}\textbf{Total}}                                                                                       & 961                                & 2341                              & 3302                                                                                 & 70.90                               & 29.10    \\ \bottomrule                           
\end{tabular}
\vspace{-15pt}
\end{table}

\begin{figure}[t]
\centering
\subfloat[DNN]{
\includegraphics[width=.48\linewidth]{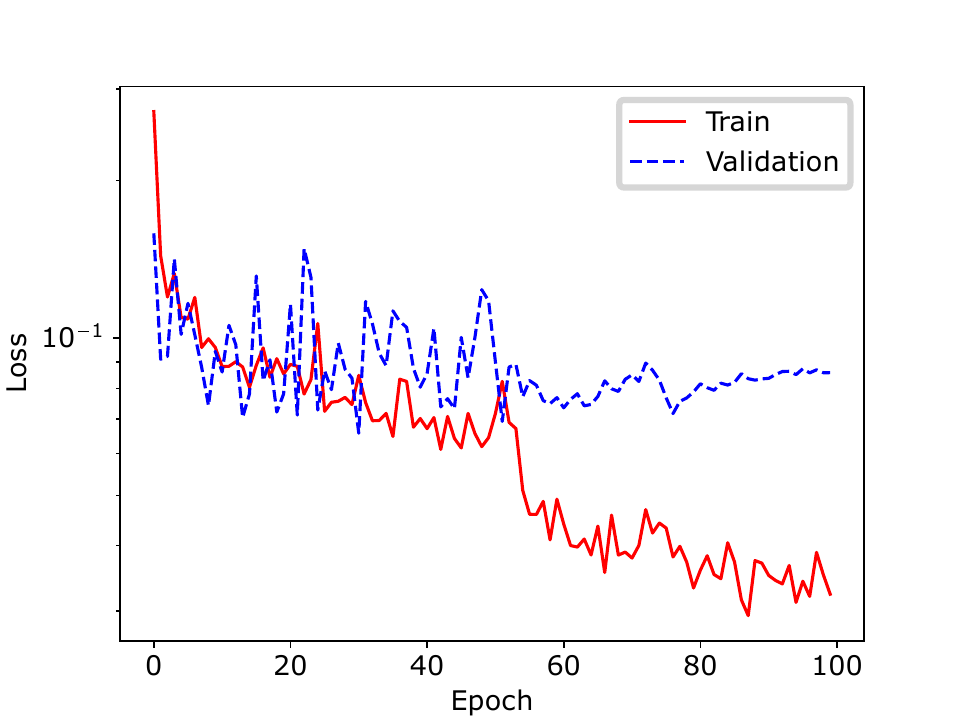}
}
\subfloat[QNN]{    
\includegraphics[width=.48\linewidth]{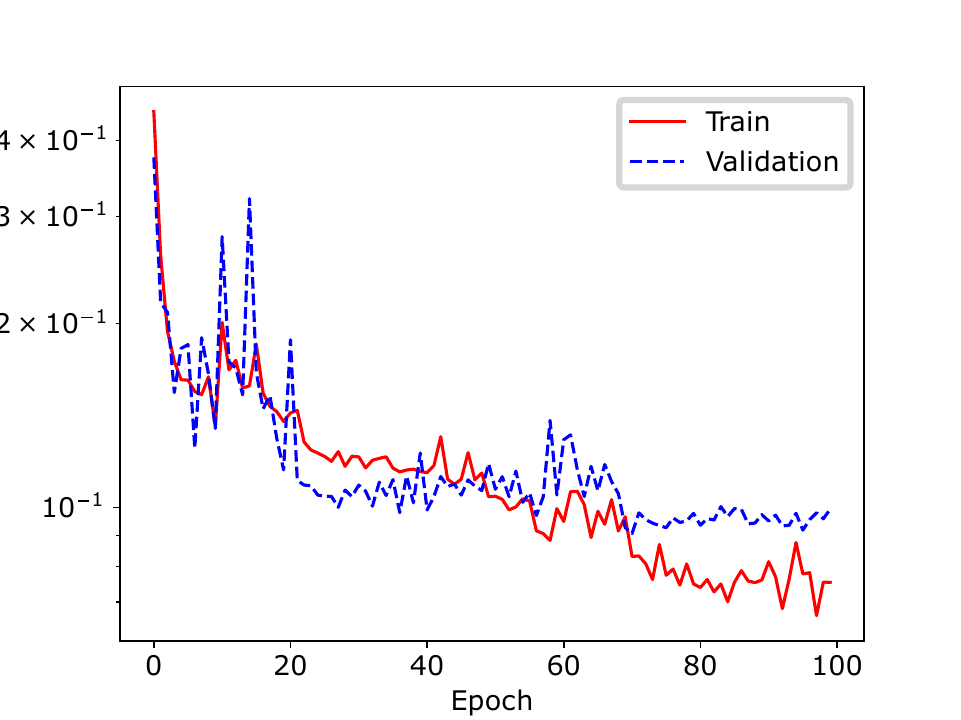}
}
\caption{ML model training: Cross-entropy loss over epoch.}
\label{fig:model_training}
\end{figure}

\begin{figure}[t]
\centering
\subfloat[ROC for competing models]{
\includegraphics[width=.48\linewidth]{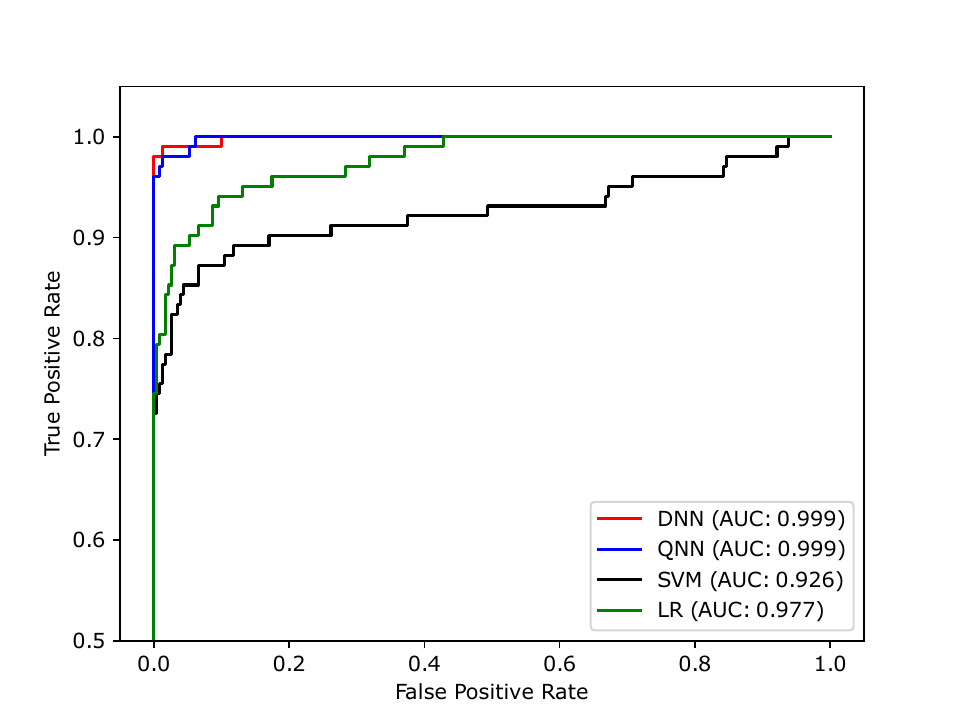}
}
\subfloat[ROC-AUC vs Training size]{    
\includegraphics[width=.48\linewidth]{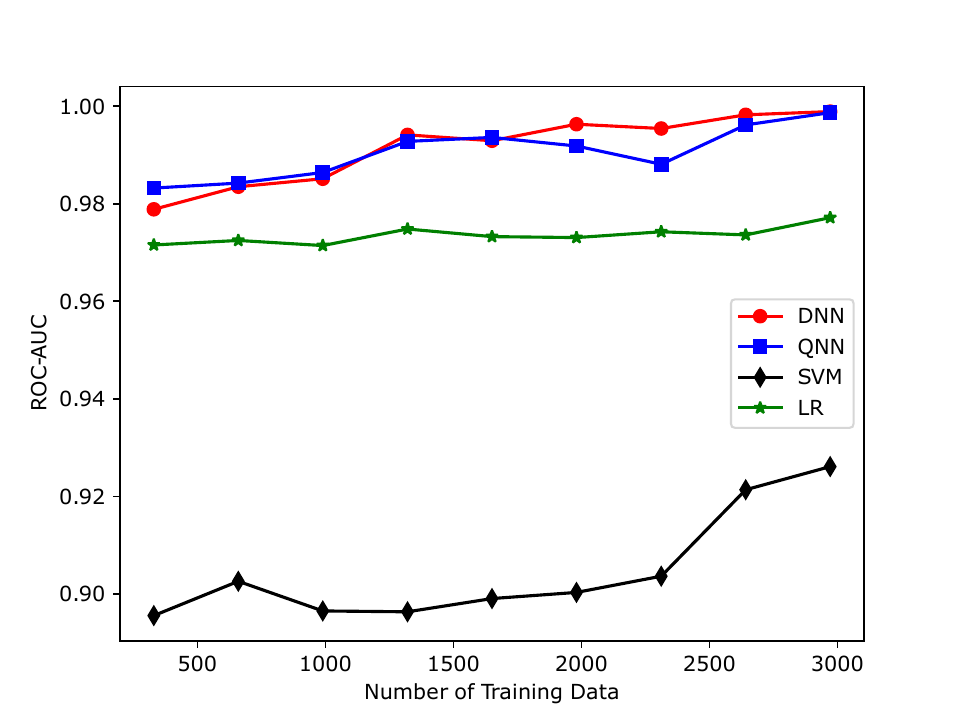}
}

\caption{ML model Accuracy of Training data size}
\vspace{-15pt}
\label{fig:ROC-AUC}
\end{figure}

\subsection{Training and Validation}
We generate 3302 convolution design points with BO, and 961 of them are synthesizable within the given time budget. Each sample has five independent variables, one dependent variable, and a kernel identifier. We use all samples to train classification models and synthesizable samples to train regression models. Classification models predict the sample outcome and are used as early failure prediction models. Regression models predict FPGA resource usage. We show the model training process over 100 epochs and the corresponding loss in Fig.~\ref{fig:model_training}. Our models converge quickly on the training data. Fig.~\ref{fig:ROC-AUC} (b) demonstrates that our models can learn from a small number of training samples and achieve high accuracy on the test data.

\begin{figure}[t]
\centering
\subfloat[LUT prediction]{
\includegraphics[width=.48\linewidth]{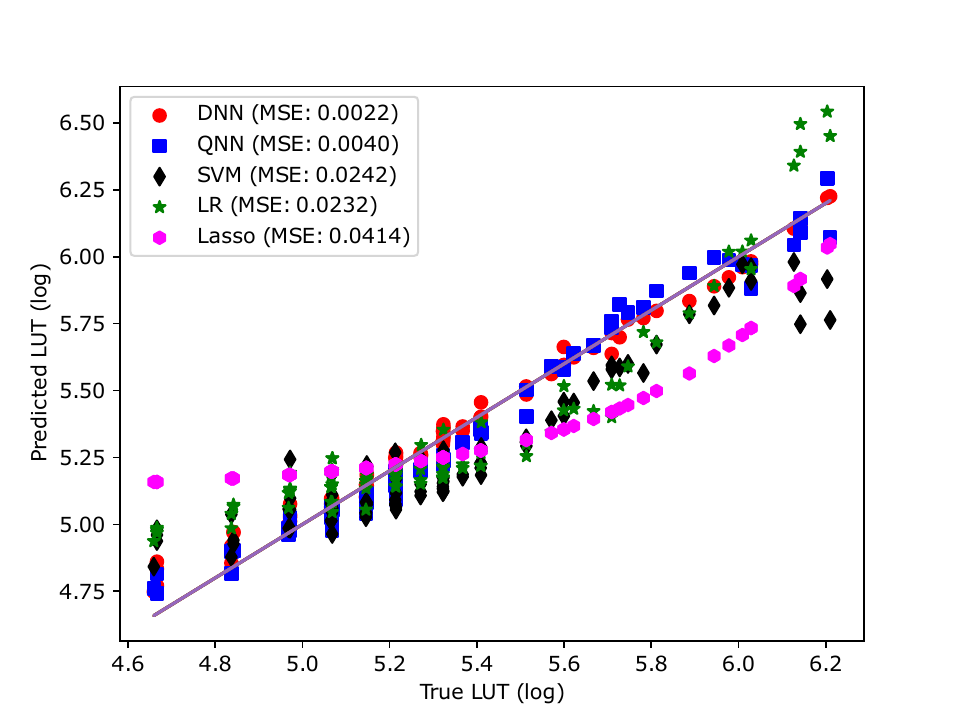}
}
\subfloat[Pareto optimization]{    
\includegraphics[width=.48\linewidth]{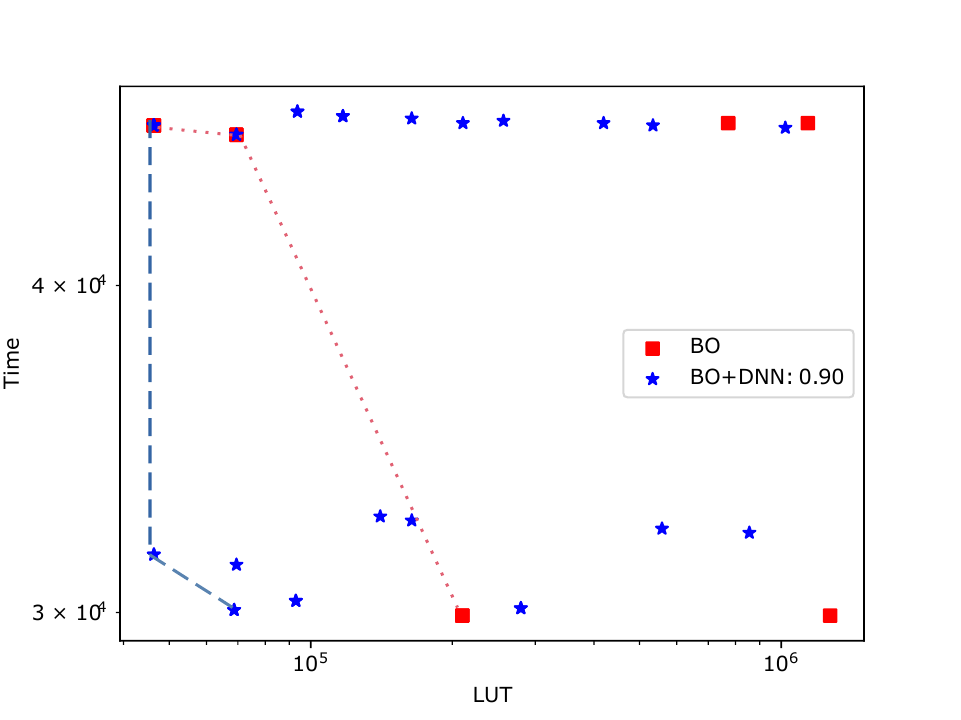}
}

\caption{FPGA synthesis results using different models.}
\label{fig:lut_pareto}
\end{figure}

Proposed models are validated under various conditions. The results show high true positive rates in the ROC curve, as demonstrated in Fig.~\ref{fig:ROC-AUC} (a). The models' robustness and generalization capabilities are also confirmed. They still achieve high true positive rates even when trained on only 5\% of the samples. Proposed DNN and QNN models outperform classical regression methods, as shown in Fig.~\ref{fig:lut_pareto} (a). Regarding Pareto fronts, \autohls outperforms BO as highlighted in the blue dotted line in Fig.~\ref{fig:lut_pareto} (b).

We evaluate the effectiveness of the proposed early failure prediction model by running a pragma parameter exploration for the APoT kernel. The estimated time for each design point synthesis is ten minutes. Table~\ref{tab:speedup} shows the results for different threshold values $\tau$, demonstrating a speedup in synthesizable design exploration time ranging from 15 to 74 times faster when using the failure prediction model.

\begin{table}[!t]
\caption{AutoHLS parameter search with failure prediction.}
\label{tab:speedup}
\centering
\begin{tabular}{llllllll}
\toprule
\textbf{$\tau$} &
  \multicolumn{1}{c}{\textbf{samples}} &
  \multicolumn{1}{c}{\textbf{TP}} &
  \multicolumn{1}{c}{\textbf{FP}} &
  \textbf{\%TP} &
  \multicolumn{1}{c}{\textbf{\begin{tabular}[c]{@{}c@{}}BO\\ hrs.\end{tabular}}} &
  \multicolumn{1}{c}{\textbf{\begin{tabular}[c]{@{}c@{}}AutoHLS\\ hrs\end{tabular}}} &
  \multicolumn{1}{c}{\textbf{Speedup}} \\ \midrule
0.95 & 2000 & 48 & 4 & 2.5  & $\sim$333 & $\sim$8   & $\sim$38 \\
0.85 & 2000 & 23 & 4 & 1.15 & $\sim$333 & $\sim$4.5 & $\sim$\textbf{74} \\
0.75 & 200  & 14 & 1 & 7.0    & $\sim$33  & $\sim$2.5 & $\sim$14 \\
\bottomrule
\end{tabular}
\vspace{-15pt}
\end{table}

\subsection{Discussion}
An important concern regarding \autohls is its generality. Future research can experiment with unseen designs to evaluate the generalizability of this framework. Improvements can also be made by leveraging the vast amount of open-source FPGA synthesis data available in DB4HLS~\cite{DB4HLS}, which contains more than 100,000 design points. Our experiments with the CNN kernel demonstrate \autohls's efficacy, even with an imbalanced training set. The low false positive rate achieved by \autohls indicates that the machine learning models can learn effectively. We consider synthesizability within a given time budget and note that early failure prediction could be possible for other metrics, such as DSP and clock cycle numbers. Due to the nature of HLS synthesis data, \autohls can learn from a small number of training data. Finally, we suggest exploring multi-objective reinforcement learning methods to enhance the robustness of this framework.

\section{Conclusion}
\label{sec:conclusion}
This paper presents \autohls, a framework for accelerating DSE for HLS using DNN/QNN-enabled multi-objective BO. It addresses the shortcomings of BO in HLS optimization. Furthermore, it provides resource prediction mechanisms and faster exploration of the Pareto front. It demonstrates the effectiveness of this framework in achieving specific design goals through accelerated DSE and kernel operation transformation. Our experiments significantly speed up finding optimal FPGA design parameters for the CNN kernel.

\bibliographystyle{unsrt}
\bibliography{ref/refs}

\end{document}